\documentclass{article} % For LaTeX2e
\usepackage{preprint,times}

\usepackage{amsmath,amsfonts,bm}

\def\eqref#1{equation~\ref{#1}}
\def\1{\bm{1}}

\DeclareMathAlphabet{\mathsfit}{\encodingdefault}{\sfdefault}{m}{sl}
\SetMathAlphabet{\mathsfit}{bold}{\encodingdefault}{\sfdefault}{bx}{n}

\usepackage{hyperref}
\hypersetup{hidelinks}
\usepackage{url}
\usepackage{multirow}
\usepackage{booktabs} % for professional tables
\usepackage[table]{xcolor}
\usepackage{graphicx}
\usepackage{caption}
\usepackage{float}
\usepackage{placeins}
\usepackage{etoc}
\usepackage{xurl}
\usepackage{xcolor}
\usepackage[most]{tcolorbox}
\usepackage{booktabs}
\usepackage[table]{xcolor}
\usepackage{arydshln} % provides \hdashline
\etocdepthtag.toc{mtchapter}
\etocsettagdepth{mtchapter}{subsubsection}
\etocsettagdepth{mtappendix}{none}

\newcommand{\modelicon}[1]{%
  \makebox[1.35em][c]{%
    \raisebox{-0.15em}{%
      \includegraphics[width=1.05em,height=1.05em,keepaspectratio]{fig/icon/#1}%
    }%
  }\hspace{0.25em}%
}
\newcommand{\linkicon}[1]{%
  \raisebox{-0.2em}{\includegraphics[height=1.25em,keepaspectratio]{fig/icon/#1.png}}\hspace{0.4em}%
}
\newlength\savewidth
\newcommand{\tablestyle}[2]{\setlength{\tabcolsep}{#1}\renewcommand{\arraystretch}{#2}\centering\footnotesize}
\renewcommand{\paragraph}[1]{\vspace{1.25mm}\noindent\textbf{#1}}
\newcommand\blfootnote[1]{%
  \begingroup
  \renewcommand\thefootnote{}%
  \NoHyper\footnotetext{\hspace*{-1.8em}\ignorespaces#1}\endNoHyper%
  \endgroup
}

\title{Doing More with Less Tokens: Hierarchical Reinforcement Learning for Efficient Coding Agents}

\author{Haobin Li \and Liang Jiang \and Zhenyu Huang \and Mouxing Yang \and Xi Peng}
\hypersetup{
  pdftitle={Doing More with Less Tokens: Hierarchical Reinforcement Learning for Efficient Coding Agents},
  pdfauthor={Haobin Li, Liang Jiang, Zhenyu Huang, Mouxing Yang, Xi Peng}
}
\newcommand{\heromark}[1]{\raisebox{0.7ex}{\fontsize{7}{8}\selectfont\mdseries #1}}
\newcommand{\heroauthor}[2]{\textbf{#1}\heromark{#2}}
\newcommand{\herohfurl}{https://huggingface.co/collections/XLearning-SCU/hero}
\newcommand{\herohflink}{%
  \ifx\herohfurl\empty
    \linkicon{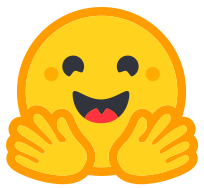}\textbf{Hugging Face}%
  \else
    \linkicon{huggingface}\href{\herohfurl}{\textbf{Hugging Face}}%
  \fi
}
\newcommand{\makeherofront}{%
\noindent\begin{minipage}{\textwidth}
  \centering
  \includegraphics[width=0.35\textwidth]{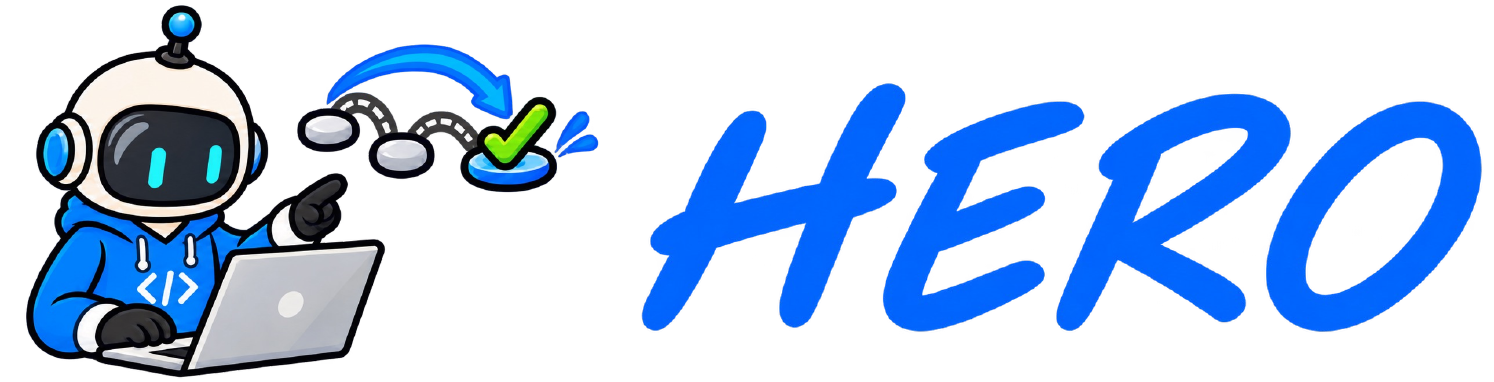}\par
  \vspace{1mm}
  {\fontsize{16}{19}\selectfont\bfseries
    Doing More with Less Tokens: Hierarchical Reinforcement Learning for Efficient Coding Agents\par
  }
  \vspace{3.5mm}
  {\fontsize{11}{15}\selectfont
    \heroauthor{Haobin Li}{1}\quad
    \heroauthor{Liang Jiang}{2}\par
    \vspace{1mm}
    \heroauthor{Zhenyu Huang}{1}\quad
    \heroauthor{Mouxing Yang}{1}\quad
    \heroauthor{Xi Peng}{1}\par
  }
  \vspace{2mm}
  {\fontsize{10}{14}\selectfont
    \heromark{1}Sichuan University\hspace{1.2em}
    \heromark{2}Independent Researchers\par
  }
  \vspace{4mm}
  \begin{minipage}{\linewidth}
  \linespread{1.1}\fontsize{10}{12}\selectfont
  \setlength{\parindent}{0pt}
Recently, coding agents have emerged as a dominant paradigm for real-world software engineering (SWE) scenarios, which solve complex tasks through multi-turn interactions with development environments.
However, frequent interactions with environments would inevitably introduce substantial token overhead, leading to high usage costs and
latency.
Although recent studies have explored reducing token usage by context manipulation and interaction limits at inference time, these approaches focus on improving token efficiency while overlooking the risk of discarding task-relevant information, thus struggling to balance the trade-off between resolution rate and token efficiency.
In this paper, we study a more general paradigm without suffering from the limitation, \textit{i.e.}, training token-efficient coding agents with promising resolution performance, which is a highly-practical yet less-explored problem.
To this end, we reveal two core observations in SWE scenarios: i) \textit{Efficiency Variation}: successful resolution could be achieved with fewer tokens; ii) \textit{Entropy Correlation}: unproductive behaviors are associated with turn-level entropy.
Motivated by observations, we propose a novel reinforcement learning framework, dubbed HERO.
Specifically, HERO prioritizes task resolution over token efficiency during policy optimization and encourages efficient reasoning patterns at both trajectory and turn levels.
Extensive experiments on SWE-bench Verified and SWE-bench Multilingual demonstrate that HERO achieves a favorable trade-off between resolution rate and token efficiency compared with state-of-the-art coding agents and reinforcement learning methods.
  \end{minipage}
  \par\vspace{3mm}
  {\normalfont
    \linkicon{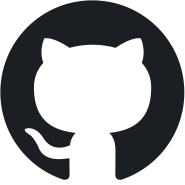}\href{https://github.com/XLearning-SCU/HERO}{\textbf{GitHub}}
    \qquad\qquad
    \herohflink
    \par
  }
\end{minipage}%
}

\begin{document}

\thispagestyle{plain}
\vspace*{-2.5em}
\makeherofront
\vspace{4mm}
\blfootnote{%
  \footnotesize
  Emails: \{haobinli.gm, zyhuang.gm, yangmouxing, pengx.gm\}@gmail.com
}

\section{Introduction}
Recently, coding agents~\citep{coding_agent1,coding_agent2,coding_agent3} powered by large language models (LLMs)~\citep{llm1,llm2} have emerged as a dominant paradigm for real-world software engineering (SWE)~\citep{swe1,swe2,active-swe} scenarios, demonstrating strong performance across various complex tasks, including but not limited to repository-level bug fixing, feature implementation, and end-to-end software development.
Specifically, coding agents first interact with environments over multiple turns through different tools, \textit{e.g.}, search, editing, and command-line tools.
By incorporating tool feedback from environments, coding agents could continually accumulate task-relevant information and thus facilitate the resolution of complex SWE tasks.

Despite the promising performance of existing coding agents, the frequent interactions with environments would inevitably introduce substantial token overhead, leading to high usage costs and latency~\citep{swe_eff}.
To remedy this, recent studies have explored pruning~\citep{swe_pruner_pro} or compressing~\citep{compact1,compact2} the input context and limiting interaction turns~\citep{turn_limit} at inference time. 
However, such inference-time approaches focus on improving token efficiency while overlooking the risk of discarding task-relevant information, thus struggling to balance the trade-off between resolution performance and token efficiency.
% However, the former approaches might overwrite context and thus undermine prefix-cache reuse, while the latter approaches heavily relies on manually predefined heuristics that are difficult to adapt to diverse tasks.
% More importantly, such inference-time designs improve efficiency at the risk of discarding task-relevant information, thus fail to balance the trade-off between resolution performance and token efficiency.
Different from the above inference-time designs, an alternative paradigm is to internalize both effective and efficient reasoning patterns into the agent policy through training-time designs, thereby reducing token usage without sacrificing the solution capability.

To learn effective and efficient reasoning patterns, the most promising solution might be token-efficient reinforcement learning (RL)~\citep{efficient_rl1,LASER,SlimSearcher}, which encourages task success and penalizes token-intensive trajectories during policy optimization.
However, existing methods are specifically designed for non-interactive reasoning or simple agentic tasks, which are inadequate for long-horizon SWE tasks due to the following two reasons.
On the one hand, it remains under-explored whether coding agents could achieve a desirable trade-off between task success and token efficiency.
On the other hand, simply penalizing token-intensive trajectories would discourage all interaction turns, including necessary reasoning steps, thereby hindering the acquisition of task-relevant information.

\begin{figure}[t]
    \centering
    \includegraphics[width=0.95\linewidth]{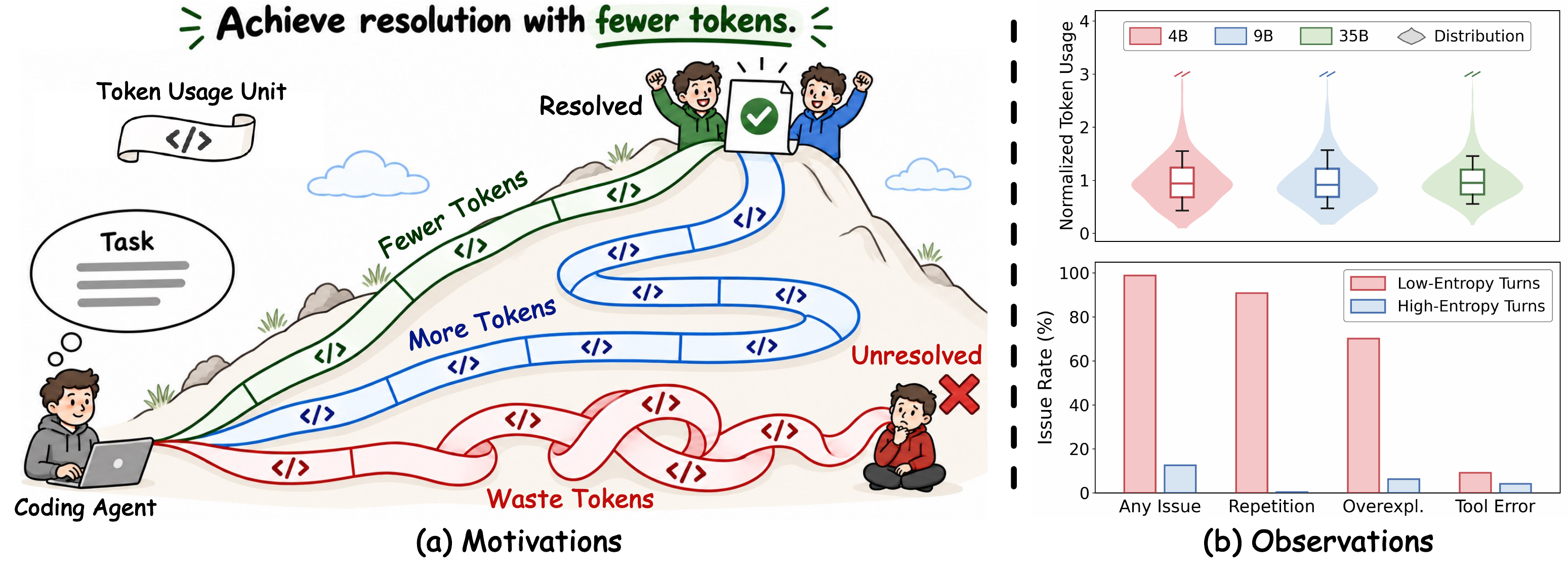}
    \caption{(a) \textbf{Motivations}: for a given SWE task, the coding agent could generate multiple reasoning trajectories with different token usages and resolution outcomes. Ideally, the goal of token-efficient coding agents is to achieve resolution with fewer tokens. (b) \textbf{Observations}: we compare token usage among successful trajectories within each task and analyze the correlation between turn-level entropy and unproductive behaviors. Note that token usage is normalized by the per-task mean over successful trajectories. As a result, one could observe that i) successful trajectories exhibit substantial variation in token usage, and ii) low-entropy turns are more prone to reasoning issues, \textit{i.e.}, repeated tool calls, excessive exploration, and erroneous commands. Together, these observations provide guidance for learning effective and efficient reasoning patterns.}
    \label{fig: motivation}
\end{figure}

To specifically develop a token-efficient RL method for training coding agents, we present two core observations as illustrated in Fig.~\ref{fig: motivation}.
To be specific, 
i) \textit{Efficiency Variation}: for a given task, it is possible to achieve successful resolution through a more efficient reasoning path with fewer tokens, paving the way to develop token-efficient coding agents;
ii) \textit{Entropy Correlation}: unproductive turns could be identified based on the turn-level entropy, providing a promising avenue to learn fine-grained efficient patterns through penalizing inefficient turns.

Based on the above observations, we develop token-efficient coding agents by learning efficient reasoning patterns while preserving resolution capability.
To this end, we propose a novel training framework, dubbed HiErarchical ReinfOrcement learning (HERO), which couples a hierarchical policy optimization framework with multi-granularity credit assignment.
In brief, HERO consistently prioritizes task resolution over token efficiency and activates efficiency optimization once sufficient problem-solving capability is acquired.
Thanks to such a hierarchical optimization objective, coding agents could achieve a favorable trade-off between resolution rate and token efficiency.
Moreover, HERO adopts coarse-grained trajectory credit to encourage concise reasoning paths and fine-grained turn credit to penalize unproductive turns, thereby learning efficient reasoning patterns at multiple granularities.
Experiments demonstrate that, using only 640 training tasks, HERO improves the average resolution rate from 37.4\% to 42.2\% across Qwen3.5 variants on SWE-bench series. 
Compared with GRPO, HERO yields better average improvement in resolution rate with 39.8\% less token usage.

In summary, the major contributions and novelties of this
work are given as follows.
\begin{itemize}
    \item Different from existing inference-time approaches, we aim to internalize both effective and efficient reasoning patterns into the agent policy through training-time designs, thus achieving token-efficient coding agents without sacrificing resolution performance.
    \item We reveal two core observations regarding token-efficient coding agents in SWE scenarios, which not only demonstrate the potential to achieve a good trade-off between task success and token efficiency but also pave an avenue toward better efficiency through fine-grained optimization.
    \item We propose a novel method termed HERO, which couples resolution-prioritized policy optimization with multi-granularity credit assignment at trajectory and turn levels, thereby improving token efficiency while preserving resolution performance.
    \item Extensive experiments on SWE-bench Verified and SWE-bench Multilingual demonstrate that the proposed HERO achieves a favorable trade-off between resolution rate and token efficiency, significantly outperforming existing state-of-the-art coding agents and reinforcement learning methods.
\end{itemize}

\section{Related Work}
In this section, we provide a brief review of two topics highly related to this work, including coding agents and token-efficient reinforcement learning.

\subsection{Coding Agents}
Coding agents aim to solve real-world software engineering tasks by equipping LLMs with various tools to interact with environments.
Based on how LLMs are endowed with tool-use capabilities, existing approaches could be broadly grouped into two categories:
i) scaffold-based methods~\citep{coding_agent1,openhands,agentless}, which design general-purpose tool commands and interaction workflows that could be applied to most widely-used LLMs;
ii) training-based methods~\citep{swe-gym,r2e-gym,swe-master}, which enhance the ability to use the specific agent scaffold through supervised fine-tuning or reinforcement learning.

Despite their promising performance, existing coding agents focus on improving resolution rate for complex SWE tasks, while paying limited attention to the substantial token usage incurred by long-horizon interactions.
Different from them, this work aims to train token-efficient coding agents that simultaneously account for reasoning efficiency and resolution performance.

\subsection{Token-Efficient Reinforcement Learning}
Token-efficient reinforcement learning aims to reduce the token usage of LLMs through policy optimization.
According to the application scenarios, existing approaches could be broadly divided into two categories:
i) reasoning-oriented methods~\citep{efficient_rl1, gfpo, LASER}, which encourage concise chain-of-thought reasoning by adopting explicit length constraints or length-based rollout filtering;
ii) agent-oriented methods~\citep{otc,SlimSearcher}, which penalize excessive cumulative token usage over multiple turns and are primarily adopted in simple agentic tasks, \textit{e.g.}, web search and multi-hop question answering.

The major differences between existing token-efficient RL methods and this work are given below. 
On the one hand, existing methods overlook the non-negligible unproductive turns in long-horizon SWE tasks, whereas our approach could identify and discourage unproductive behaviors and thus facilitate token efficiency in multiple interactions.
On the other hand, to the best of our knowledge, our approach is one of the first works to explore simultaneously preserving resolution rate and improving token efficiency in training coding agents.

\section{Method}
In this section, we introduce HiErarchical ReinfOrcement learning (HERO) for training efficient coding agents.
In Section~\ref{sec:problem_formulation}, we present the goal of training token-efficient coding agents.
In Section\ref{sec:hpo}, we propose hierarchical policy optimization to preserve resolution performance during efficiency optimization.
In Section~\ref{sec:multi_granularity_efficiency}, we introduce multi-granularity credit assignment to learn efficient reasoning patterns at multiple granularities.

\subsection{Problem Formulation}
\label{sec:problem_formulation}

Let $x_i=(r_i,s_i)$ denote a repository-level task in SWE scenarios, where $r_i$ is the repository environment and $s_i$ is the corresponding issue report.
For a given $r_i$ with $s_i$, the goal of coding agent $\mathcal{A}$ is to interact with environment $r_i$ for $n_i$ turns and then generate solution $\mathcal{A}(r_i,s_i)$  for $s_i$.
To verify the effectiveness of the generated solution, we adopt the test-driven evaluation protocol and the corresponding resolution outcome could be derived as,
\begin{equation}
    \label{eq: resolution-outcome}
    y_i
    =
    \mathcal{P}_i
    \left(
        \mathcal{A}(r_i,s_i)
    \right)
    \in\{0,1\},
\end{equation}
where $\mathcal{P}_i(\cdot)$ indicates whether the generated solution $\mathcal{A}(r_i,s_i)$ passes all task-specific tests.
As discussed in the Introduction, beyond resolution performance, we further consider the cumulative token usage during multi-turn interactions.
For a given task $x_i$, the cumulative token usage is defined as,
\begin{equation}
    \label{eq: token-usage}
    c_i=\sum_{j=1}^{n_i}t_{i,j},
\end{equation}
where $t_{i,j}=t_{i,j}^{\mathrm{in}}+t_{i,j}^{\mathrm{out}}$ denotes the token usage at the $j$-th turn, with $t_{i,j}^{\mathrm{in}}$ and $t_{i,j}^{\mathrm{out}}$ representing the input and output tokens, respectively.

With the above formulation, we define the objective of token-efficient coding agents as maximizing the Token-efficient Resolution Score (TRS), \textit{i.e.},
\begin{equation}
    \label{eq: trs}
    \max_{\theta}\ \mathrm{TRS},
    \qquad
    \mathrm{TRS}
    =
    \frac{R}{\sqrt{1+C}},
\end{equation}
where $R=\frac{100}{N}\sum_{i=1}^{N}y_i$ and $C=\frac{1}{N}\sum_{i=1}^{N}c_i$ denote the average resolution rate and cumulative token usage over $N$ tasks, respectively.
Note that $C$ is measured in millions of tokens.
Clearly, a higher TRS generally indicates a more favorable trade-off between resolution rate and token usage.
However, existing coding agents are typically designed to improve resolution performance, while overlooking the substantial token overhead incurred by multi-turn interactions.

\begin{figure}[t]
    \centering
    \includegraphics[width=0.98\linewidth]{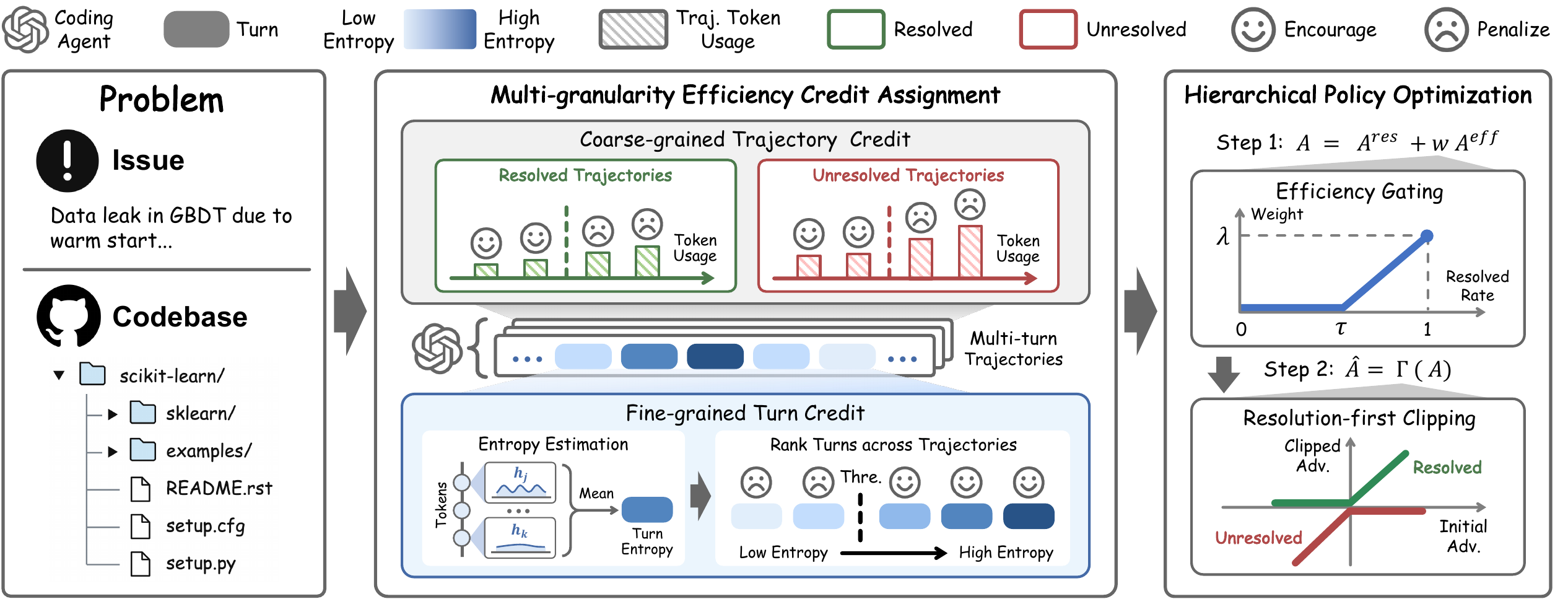}
    \caption{Overview of our method HERO. Given a swe task, the coding agent first samples multiple trajectories to form a rollout group and then computes multi-granularity advantages at trajectory and turn levels. Specifically, HERO encourages concise reasoning trajectories with the same outcome, while penalizing potentially unproductive behaviors with low turn entropy. After that, HERO adaptively activates efficiency optimization based on the resolution rate within the group and adopts resolution-first clipping to guarantee that resolved trajectories receive advantages no lower than unresolved ones.}
    \label{fig:hero-overview}
\end{figure}

As a remedy, we propose HiErarchical ReinfOrcement learning (HERO) as illustrated in Fig.~\ref{fig:hero-overview}, which couples hierarchical policy optimization over resolution and efficiency objectives with the multi-granularity efficiency credit assignment mechanism.
In the following, we will elaborate on each of them.

\subsection{Hierarchical Policy Optimization}
\label{sec:hpo}
As discussed in Introduction, existing token-efficient RL methods struggle to achieve a favorable trade-off between resolution rate and token efficiency.
To address this, we propose a novel hierarchical policy optimization (HPO) mechanism that improves resolution performance while reducing token usage, \textit{i.e.},
\begin{equation}
    \label{eq: hierarchical-rl}
\begin{gathered}
    \mathcal{L}(\theta)
    =
    -\mathbb{E}_{i\in\mathcal{G}}
    \left[
        \min
        \left(
            \rho\left(\theta\right)\widehat{A}_i,
            \operatorname{clip}
            \left(
                \rho\left(\theta\right),1-\epsilon,1+\epsilon
            \right)\widehat{A}_i
        \right)
    \right],\\
    \widehat{A}_i
    =
    \Gamma(A_i),
    \qquad
    A_i
    =
    A_i^{\mathrm{res}}
    +
    wA_i^{\mathrm{eff}},
\end{gathered}
\end{equation}
where $\mathcal{G}$ denotes the rollout group for the same task, $\rho(\theta)$ denotes the importance ratio between the current and rollout policies, $\epsilon$ indicates the clipping threshold.
In particular, $\Gamma(\cdot)$ denotes the resolution-first clipping strategy, $A_i^{\mathrm{res}}$ and $A_i^{\mathrm{eff}}$ represent the resolution and efficiency advantages of the $i$-th rollout in $\mathcal{G}$, respectively.
In the implementation, we derive $A_i^{\mathrm{res}}$ using the standard group-relative advantage estimator in GRPO~\citep{grpo}, with the binary resolution outcome $0$ or $1$ as the reward.
More specifically, we adopt an efficiency gating mechanism to balance the trade-off between resolution performance and token efficiency, \textit{i.e.},
\begin{equation}
    \label{eq: adaptive-weight}
    w
    =
    \lambda
    \operatorname{clip}
    \left(
        \frac{p-\tau}{1-\tau},
        0,
        1
    \right), \quad 
    p
    =
    \frac{1}{|\mathcal{G}|}
    \sum_{i\in\mathcal{G}}y_i.
\end{equation}
where $\lambda$ controls the strength of efficiency optimization, $p$ indicates the average resolution rate within the group $\mathcal{G}$, and $\tau$ denotes the threshold that determines whether efficiency optimization is activated.
Such a mechanism encourages coding agents to learn efficient reasoning patterns only when they demonstrate sufficient capability to resolve the given task, thereby prioritizing task resolution before optimizing token efficiency.

Although the efficiency gating mechanism poses constraints on efficiency optimization, the efficiency signal might override the resolution signal, causing unresolved trajectories to receive higher advantages than resolved ones.
To avoid this, we employ the resolution-first clipping strategy as follows,
\begin{equation}
    \label{eq: resolution-clipping}
    \Gamma(A_i)
    =
    \begin{cases}
        \max(A_i,0), & y_i=1\\
        \min(A_i,0), & y_i=0
    \end{cases}.
\end{equation}
Thanks to the resolution-first clipping strategy, advantages are non-negative for resolved trajectories and non-positive for unresolved ones, thereby further preserving the priority of task resolution during policy optimization.

\subsection{Multi-granularity Efficiency Credit Assignment}
\label{sec:multi_granularity_efficiency}
To prevent inefficient behaviors from incurring substantial token overhead, we introduce a multi-granularity credit assignment mechanism, which assigns efficiency credit at both the coarse-grained trajectory level and the fine-grained turn level.
Specifically, the efficiency advantage is derived as follows,
\begin{equation}
    \label{eq: efficiency-advantage}
    A_{i,j}^{\mathrm{eff}}
    =
    A_i^{\mathrm{coarse}}
    +
    A_{i,j}^{\mathrm{fine}},
\end{equation}
where $A_i^{\mathrm{coarse}}$ denotes the coarse-grained advantage of the $i$-th rollout, $A_{i,j}^{\mathrm{fine}}$ indicates the fine-grained advantage of $j$-th turn in the corresponding rollout.
In the following, we will elaborate on the two advantages in detail.

\subsubsection{Coarse-grained Trajectory Advantage}
To facilitate the learning of efficient reasoning patterns, a straightforward approach is to penalize token-intensive trajectories while encouraging concise ones.
However, recent studies~\citep{swe_eff,Code_agent_behaviour} have demonstrated that unresolved trajectories typically consume more tokens than resolved ones.
As a result, simply rewarding trajectories with lower token usage might conflate token efficiency with resolution rate rather than encouraging more efficient reasoning patterns.

To remedy this, we adopt the resolution-calibrated credit assignment strategy as follows,
\begin{equation}
    \label{eq: coarse-advantage}
\begin{gathered}
    A_i^{\mathrm{coarse}}
    =
    \frac{
        u_i-\bar{u}
    }{
        \sigma_u
    },
    \\[3pt]
    u_i
    =
    1-\tanh
    \left(
        \max
        \left(
            \frac{c_i}{B^{y_i}}-1,\,0
        \right)
    \right),
    \qquad
    B^{y_i}
    =
    \operatorname{Median}
    \left\{
        c_k\mid k\in\mathcal{G},\ y_k=y_i
    \right\},
\end{gathered}
\end{equation}
where $u_i$ denotes the efficiency reward for the $i$-th rollout, $\operatorname{Median}\{\cdot\}$ indicates the median token usage among rollouts in $\mathcal{G}$ with the same resolution outcome, and $\bar{u}$ and $\sigma_u$ represent the mean and standard deviation of the efficiency rewards within $\mathcal{G}$, respectively.
Intuitively, efficiency rewards are calibrated within each outcome subset, thus avoiding the conflation of token efficiency with resolution performance.
After that, the calibrated rewards are standardized to obtain trajectory-level advantages, facilitating the learning of efficient reasoning patterns under the same resolution outcome.

\subsubsection{Fine-grained Turn Advantage}
Although the proposed coarse-grained advantage could reduce token usage to some extent, the trajectory-level signal treats all interaction turns equally, which overlooks that the unproductive turns are the primary source of substantial token overhead.

As discussed in the Introduction, we reveal that low-entropy turns are more likely to exhibit unproductive behaviors.
Motivated by this, we propose the following entropy-aware credit assignment mechanism to avoid inefficient behaviors, \textit{i.e.},
\begin{equation}
    \label{eq: fine-advantage}
\begin{gathered}
    A_{i,j}^{\mathrm{fine}}
    =
    \begin{cases}
        -\tanh
        \left(
            1-\dfrac{h_{i,j}}{q}
        \right),
        & h_{i,j}<q\\[4pt]
        0,
        & h_{i,j}\ge q
    \end{cases},
    \\[3pt]
    q
    =
    \operatorname{Percentile}_{0.2}
    \left(
        \left\{
            h_{i,j}
            \mid
            i\in\mathcal{G},\
            j\in\{1,\ldots,n_i\}
        \right\}
    \right),
\end{gathered}
\end{equation}
where $h_{i,j}$ denotes the average token entropy at the $j$-th turn of the $i$-th rollout.
Following~\citet{20_80}, $\operatorname{Percentile}_{0.2}(\cdot)$ returns the entropy threshold corresponding to the lowest $0.2$ fraction of turns within the group.
Such a design would penalize low-entropy turns potentially associated with inefficient behaviors, thereby reducing the token overhead incurred by unproductive interactions.

\section{Experiments}
\label{sec:experiments}
In this section, we conduct extensive experiments on widely-used SWE benchmarks to verify the effectiveness of the proposed HERO. Due to space limitations, we present more experiments in Appendix~\ref{sec: additional-experiments}.

\subsection{Experiment Configurations}
\textbf{Benchmarks. } To comprehensively evaluate the effectiveness of HERO, we evaluate our method on the most widely-used benchmarks in SWE scenarios, \textit{i.e.}, SWE-bench Verified and SWE-bench Multilingual~\citep{swebench}.
Specifically, SWE-bench Verified contains 500 high-quality issue-solving tasks constructed from real-world GitHub issues, covering bug fixing, feature implementation, code refactoring, and so on.
While SWE-bench Multilingual extends the evaluation beyond Python, comprising 300 SWE tasks across nine programming languages.

\textbf{Implementation Details.}
We evaluate our proposed HERO using the widely-used SOTA model Qwen3.5~\citep{qwen3} and conduct comprehensive experiments on three variants with different model sizes, \textit{i.e.}, Qwen3.5-4B, Qwen3.5-9B, and Qwen3.5-35B-A3B.
For a fair comparison, we adopt the most widely-acknowledged scaffold, \textit{i.e.}, Claude Code, as the unified framework for evaluating all coding agents and RL methods.
As for comparisons between RL methods, we use the same training set of 640 multilingual SWE tasks sampled from SWE-Gym~\citep{swe-gym}, Multi-SWE-bench~\citep{multiswebench}, and SWE-rebench~\citep{swerebench}.
See Appendix~\ref{sec:training-data-details} for more details.
Regarding hyperparameter settings, the trade-off coefficient $\lambda$ and threshold $\tau$ in Eq.~\ref{eq: adaptive-weight} are set to $0.3$ and $0.6$, respectively.
We repeat all experiments three times with different random seeds and report the mean results.

\begin{table}[t]
\centering
\caption{Comparisons with state-of-the-art methods on SWE-bench Verified benchmark. Token usage is reported in millions (M).}
\label{tab:swe_verified}
\tablestyle{11pt}{1.3}
\begin{tabular}{lccc}
\toprule
Method
& Resolve Rate (\%)
& Token Usage (M)
& TRS \\
\midrule

\multicolumn{4}{c}{\textit{\textbf{Open-Source Coding Agents}}} \\
\midrule

\modelicon{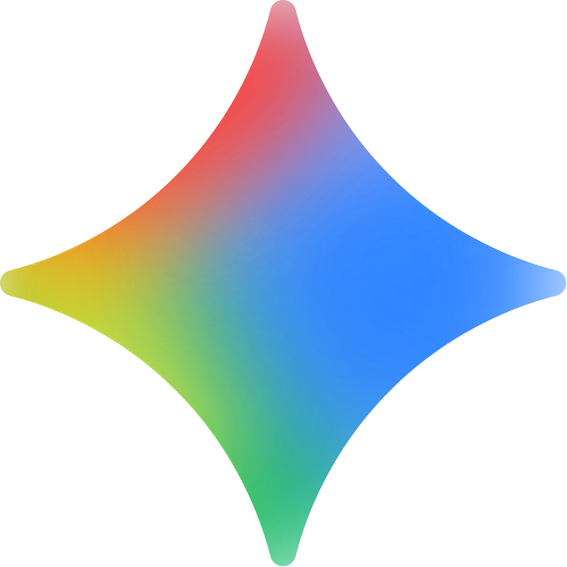}Gemma-4-26B-A4B
& 35.4
& 2.3
& 19.5 \\

\modelicon{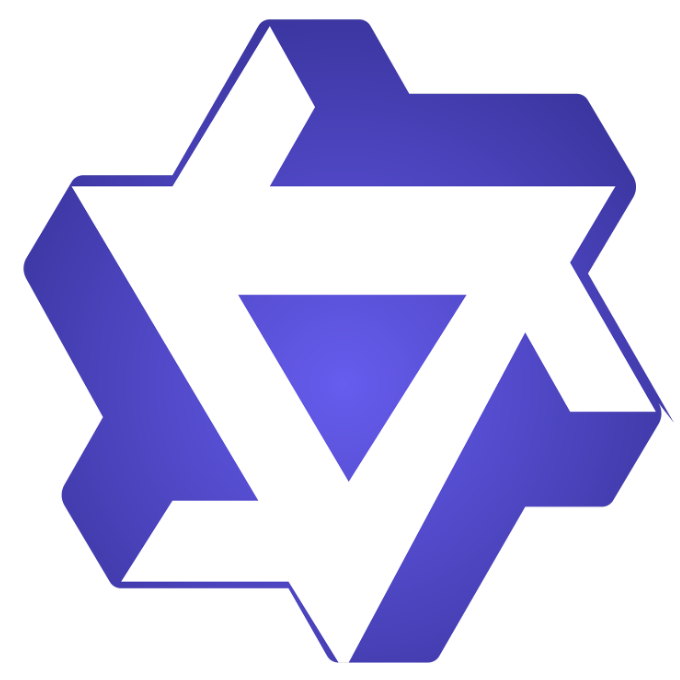}Qwen3-Coder-30B-A3B-Instruct
& 43.2
& 2.4
& 23.4 \\

\modelicon{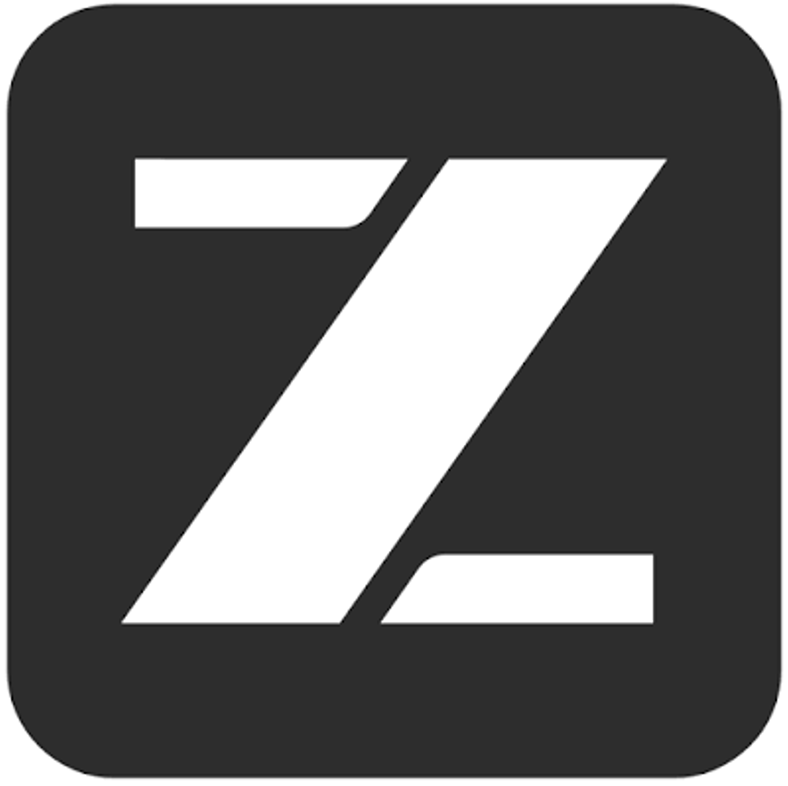}GLM-4.7-Flash-30B
& 42.6
& 4.6
& 18.0 \\

\modelicon{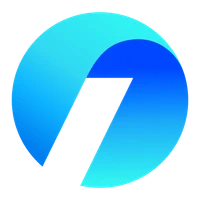}KAT-Dev-32B
& 48.4
& 2.4
& 26.2 \\

\midrule
\multicolumn{4}{c}{\textit{\textbf{Reinforcement Learning}}} \\
\midrule

\modelicon{qwen.png}Qwen3.5-4B
& 32.8
& 4.5
& 14.0 \\

\hspace{1.0em} + GRPO
& 37.4
& 6.8
& 13.4 \\

\rowcolor{pink!30}
\hspace{1.0em} + Our
& 40.0
& 4.5
& 17.1 \\

\noalign{\vskip 2pt}
\hdashline
\noalign{\vskip 2pt}

\modelicon{qwen.png}Qwen3.5-9B
& 45.0
& 4.5
& 19.2 \\

\hspace{1.0em} + GRPO
& 47.0
& 5.9
& 17.9 \\

\rowcolor{pink!30}
\hspace{1.0em} + Our
& 49.2
& 3.5
& 23.2 \\

\noalign{\vskip 2pt}
\hdashline
\noalign{\vskip 2pt}

\modelicon{qwen.png}Qwen3.5-35B-A3B
& 57.2
& 2.9
& 29.0 \\

\hspace{1.0em} + GRPO
& 59.2
& 5.2
& 23.8 \\

\rowcolor{pink!30}
\hspace{1.0em} + Our
& 60.6
& 2.7
& 31.5 \\

\bottomrule
\end{tabular}
\par\vspace{20pt}
\normalsize
\caption{Comparisons with state-of-the-art methods on SWE-bench Multilingual benchmark.}
\label{tab:swe_multilingual}
\tablestyle{11pt}{1.3}
\begin{tabular}{lccc}
\toprule
Method
& Resolve Rate (\%)
& Token Usage (M)
& TRS \\
\midrule

\multicolumn{4}{c}{\textit{\textbf{Open-Source Coding Agents}}} \\
\midrule

\modelicon{gemini.png}Gemma-4-26B-A4B
& 27.3
& 3.5
& 12.9 \\

\modelicon{qwen.png}Qwen3-Coder-30B-A3B-Instruct
& 20.6
& 2.1
& 11.7 \\

\modelicon{glm.png}GLM-4.7-Flash-30B
& 25.3
& 4.0
& 11.3 \\

\modelicon{kat.png}KAT-Dev-32B
& 27.6
& 2.3
& 15.2 \\

\midrule
\multicolumn{4}{c}{\textit{\textbf{Reinforcement Learning}}} \\
\midrule

\modelicon{qwen.png}Qwen3.5-4B
& 20.0
& 4.0
& 8.9 \\

\hspace{1.0em} + GRPO
& 22.0
& 7.6
& 7.5 \\

\rowcolor{pink!30}
\hspace{1.0em} + Our
& 24.3
& 4.5
& 10.4 \\

\noalign{\vskip 2pt}
\hdashline
\noalign{\vskip 2pt}

\modelicon{qwen.png}Qwen3.5-9B
& 29.3
& 4.3
& 12.7 \\

\hspace{1.0em} + GRPO
& 31.0
& 6.0
& 11.7 \\

\rowcolor{pink!30}
\hspace{1.0em} + Our
& 34.0
& 4.1
& 15.1 \\

\noalign{\vskip 2pt}
\hdashline
\noalign{\vskip 2pt}

\modelicon{qwen.png}Qwen3.5-35B-A3B
& 40.3
& 3.2
& 19.7 \\

\hspace{1.0em} + GRPO
& 43.0
& 5.4
& 17.0 \\

\rowcolor{pink!30}
\hspace{1.0em} + Our
& 45.0
& 2.9
& 22.8 \\

\bottomrule
\end{tabular}
\end{table}

\subsection{Comparisons with State-of-the-Art Methods}

We compare our method with the widely-used reinforcement learning method, \textit{i.e.}, GRPO~\citep{grpo}, using Qwen3.5 backbones of various model sizes.
For more results on state-of-the-art token-efficient reinforcement learning methods, please refer to Appendix~\ref{sec:training-efficient-comparison}.
Besides, we compare against four state-of-the-art coding agents, \text{i.e.}, Gemma-4-26B-A4B~\citep{gemma4}, Qwen3-Coder-30B-A3B-Instruct~\citep{qwen3}, GLM-4.7-Flash-30B~\citep{glm5}, and KAT-Dev-32B~\citep{kat}.
% For fair comparison, all experiments are conducted under the unified Claude Code scaffold.

From the results in Table~\ref{tab:swe_verified}-\ref{tab:swe_multilingual}, one could have the following observations and conclusions:
i) state-of-the-art coding agents with strong resolution performance do not necessarily exhibit high token efficiency;
ii) GRPO consistently improves resolution rates at the cost of substantially increased token usage, suggesting that GRPO could enhance issue-solving capabilities without necessarily encouraging efficient reasoning patterns;
iii) in contrast, the proposed HERO prioritizes task resolution over token efficiency while reducing the token overhead incurred by unproductive interactions, achieving desirable resolution rates and token efficiency;
iv) thanks to the proposed HERO, even small models such as Qwen3.5-4B and Qwen3.5-9B achieve competitive resolution performance against larger open-source SOTA coding agents.

% \FloatBarrier
\subsection{Ablation and Analytic Study}
In this section, we carry out a series of ablation studies and analytic experiments to investigate the effectiveness of HERO. Unless otherwise stated, experiments are conducted using Qwen3.5-4B on the SWE-bench Verified benchmark.

\begin{figure}[t]
    \centering
    \includegraphics[width=0.98\linewidth]{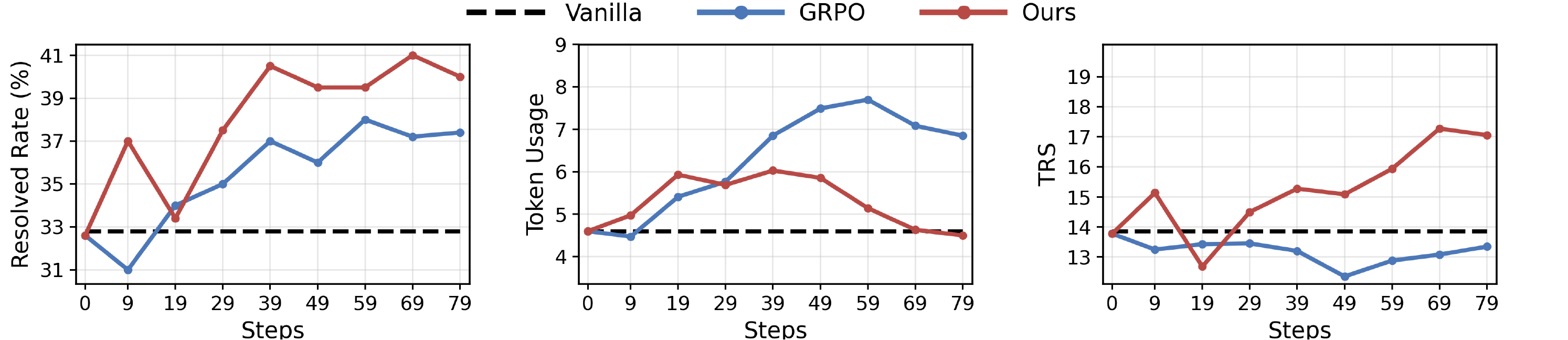}
    \caption{Dynamics of the training process.}
    \label{fig: step-metric}
\end{figure}

\textbf{Analytic Study on Training Dynamics.}
To investigate how HERO learns efficient reasoning patterns and improves resolution performance, we carry out an analytic study on the dynamics throughout the training process, including resolved rate, token usage, and TRS.
As demonstrated in Fig.~\ref{fig: step-metric}, HERO outperforms GRPO across the three metrics as training progresses, with more pronounced gains in later stages.
Besides, we observe that token usage under HERO first rises until step $40$ and then declines, whereas the resolution rate reaches a high level by this stage and remains consistently high thereafter.
In other words, such a phenomenon suggests that HERO first develops issue-solving capabilities in the early stages and subsequently reduces token usage while preserving resolution performance in the later stage, thereby achieving a favorable trade-off between the two objectives.

\begin{figure}[t]
\centering
\begin{minipage}[t]{0.45\textwidth}
    \vspace{0pt}
    \centering
    \includegraphics[width=\linewidth]{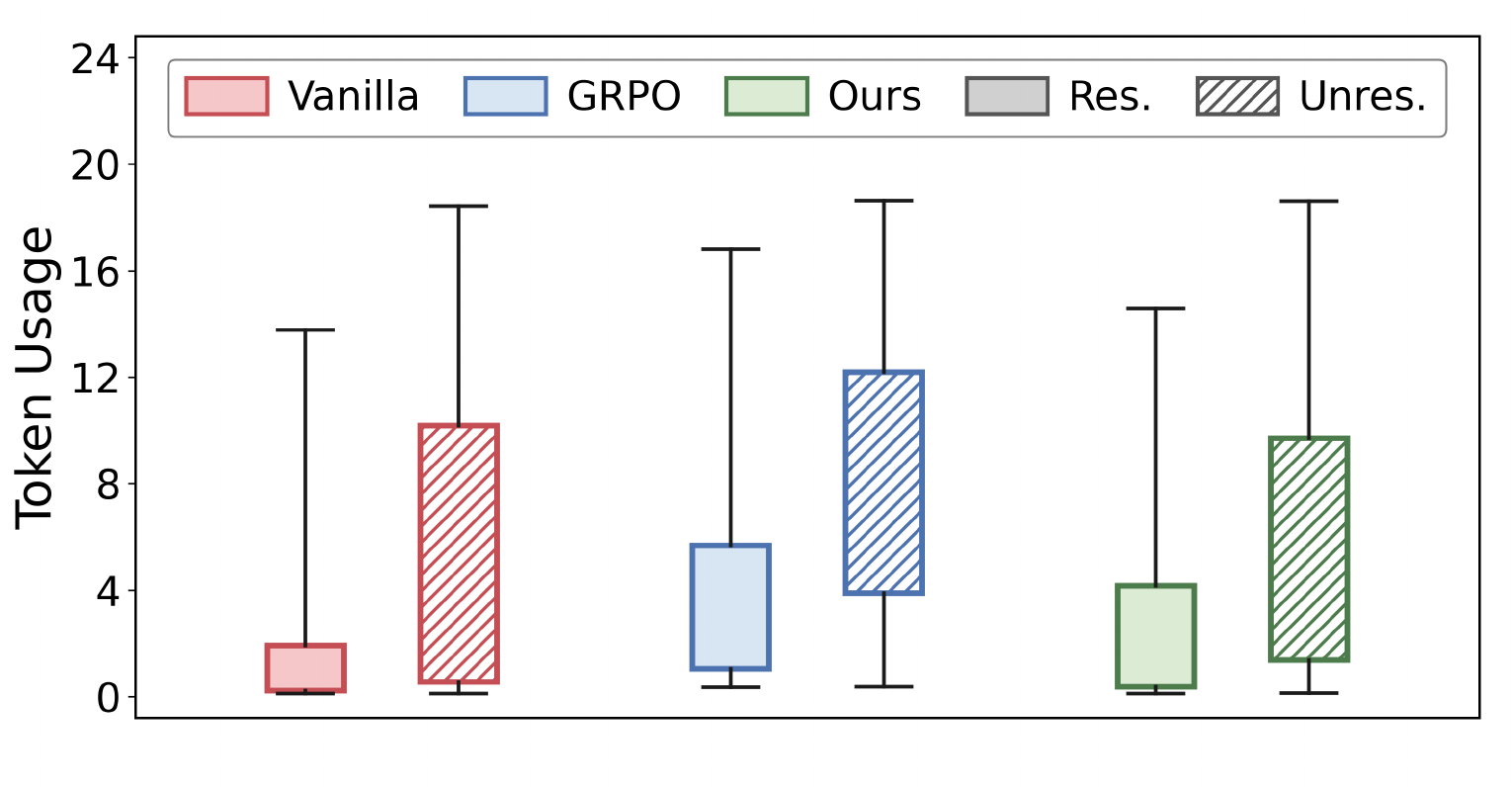}
    \captionof{figure}{Token usage distributions of resolved and unresolved trajectories.}
    \label{fig: token_comparison}
\end{minipage}
\ \ \ \ \ 
\begin{minipage}[t]{0.45\textwidth}
    \vspace{0pt}
    \centering
    \includegraphics[width=\linewidth]{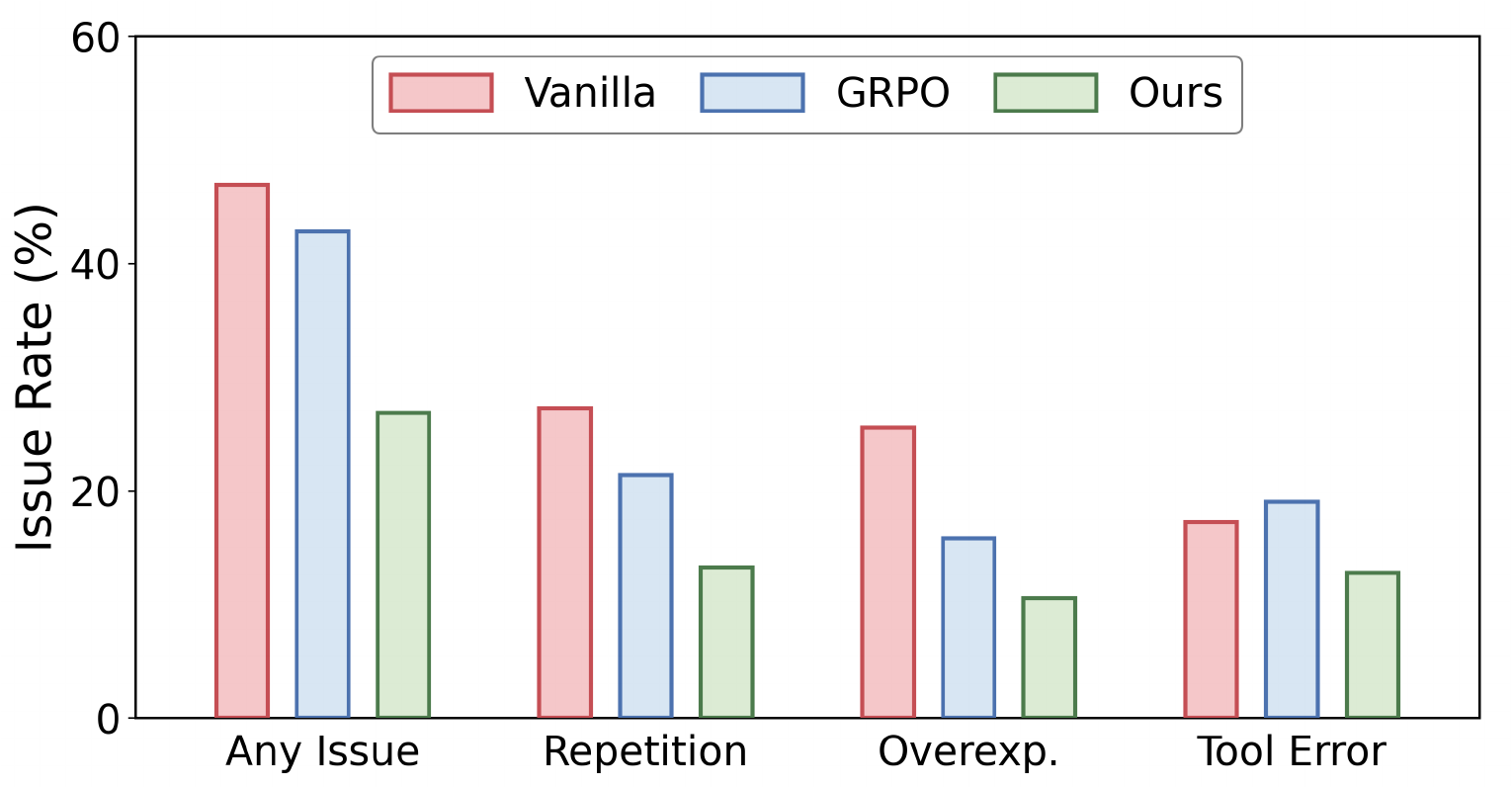}
    \captionof{figure}{Issue rate of unproductive behaviors at turn level.}
    \label{fig: turn-issue}
\end{minipage}
\end{figure}

\textbf{Analytic Study on Token Usage.}
To verify the effectiveness of HERO in improving token efficiency, we compare token usage distributions for resolved and unresolved trajectories under GRPO and the proposed HERO.
From the results in Fig.~\ref{fig: token_comparison}, HERO achieves lower token usage for both resolved and unresolved trajectories compared with GRPO, suggesting that our method could learn efficient reasoning patterns and thus achieve resolution with fewer tokens.

\textbf{Analytic Study on Unproductive Turns.}
As pointed out in Introduction, low-entropy turns are more prone to inefficient behaviors.
To verify the effectiveness of HERO in suppressing unproductive turns, we analyze the frequencies of unproductive behaviors across interaction turns.
As shown in Fig.~\ref{fig: turn-issue}, HERO achieves lower rates of repeated tool calls, overexploration, and tool errors than both the base model and GRPO, suggesting that penalizing low-entropy turns would discourage unproductive behaviors.

\begin{figure}[!t]
\centering
\begin{minipage}[t]{0.55\linewidth}
    \vspace{0pt}
    \centering
    \captionof{table}{Ablation studies on the designs in HERO, where ``w/o HPO'' denotes removing the hierarchical policy optimization.}
    \label{tab:ablation_designs}
    \begingroup
    \tablestyle{6pt}{1.15}
    \resizebox{\linewidth}{!}{%
    \begin{tabular}{lccc}
    \toprule
    Variants & Resolve Rate & Token Usage & TRS \\
    \midrule
    GRPO       & 37.4 & 6.8 & 13.4 \\   
    w/o HPO    & 29.2 & 2.1 & 16.6 \\
    w/o Coarse & 39.0 & 6.0 & 14.7 \\
    w/o Fine   & 37.8 & 4.9 & 15.6 \\
    \midrule
    Default    & 40.0 & 4.5 & 17.1 \\
    \bottomrule
    \end{tabular}%
    }
    \endgroup
\end{minipage}
\hfill
\begin{minipage}[t]{0.42\linewidth}
    \vspace{0pt}
    \centering
    \includegraphics[width=\linewidth]{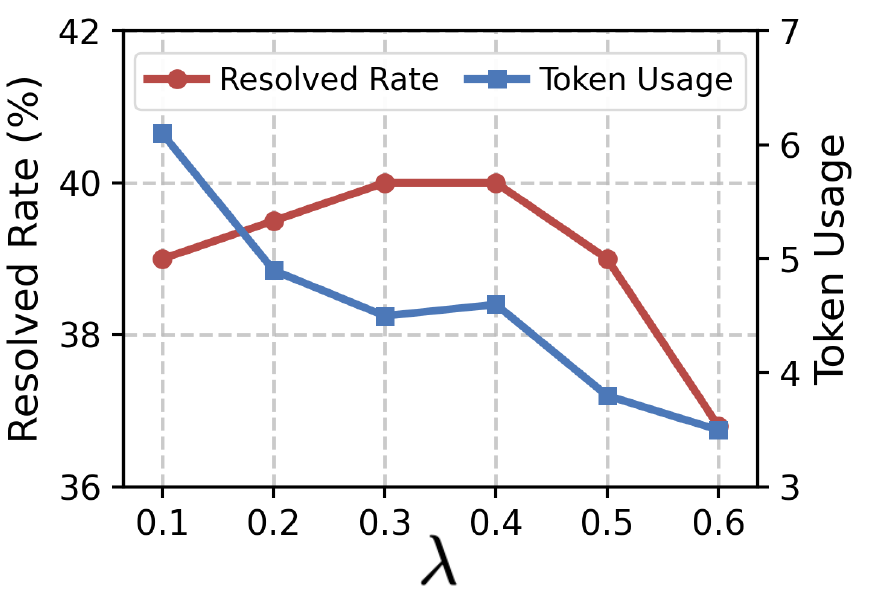}
    \vspace{-17pt}
    \captionof{figure}{Parameter analysis of eﬀiciency coefficient $\lambda$ in Eq.~\ref{eq: adaptive-weight}.}
    \label{fig: parameter-main}
\end{minipage}
\end{figure}

\textbf{Ablation Studies.}
To verify the importance of each design, we investigate the variants of HERO in Table~\ref{tab:ablation_designs}, from which one could draw the following conclusions.
On the one hand, removing hierarchical policy optimization would reduce token usage but degrade resolution performance, highlighting its role in preserving task resolution during efficiency optimization.
On the other hand, both the coarse-grained and fine-grained efficiency advantages contribute to achieving lower token usage.
Besides, we carry out parameter analysis of the efficiency coefficient $\lambda$ in Eq.~\ref{eq: adaptive-weight}.
As depicted in Fig.~\ref{fig: parameter-main}, HERO achieves stable performance improvement with low token consumption when $\lambda\in[0.2,0.4]$. See Appendix~\ref{sec: capability-threshold}-\ref{sec: entropy-percentile} for ablation studies on more parameters.

\begin{figure}[t]
    \centering
    \includegraphics[width=\linewidth]{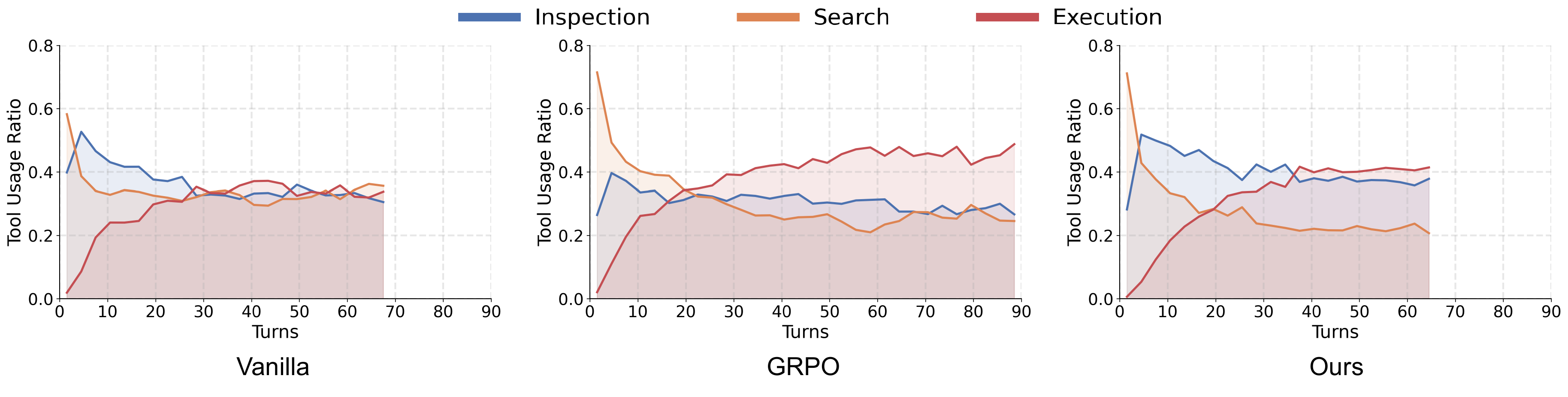}
    \caption{Tool usage across various interaction turns.}
    \label{fig: traj-behavior}
\end{figure}

\textbf{Analytic Study on Reasoning Patterns.}
To further verify that HERO learns both effective and efficient reasoning patterns, we conduct an analytical study on the tool usage across interaction turns.
From the results in Fig.~\ref{fig: traj-behavior}, we observe that HERO shifts from inspection toward execution in later turns, mirroring the reasoning progression of GRPO while requiring fewer tokens.
In other words, HERO internalizes more efficient reasoning patterns into the agent policy without undermining the necessary progression from code exploration to solution generation.

\section{Conclusion}
In this paper, we study a highly-practical yet under-explored problem, \textit{i.e.}, training token-efficient coding agents while preserving resolution performance.
To solve this problem, we propose a novel RL framework, \textit{i.e.}, HERO, which achieves efficient reasoning patterns through multi-granularity credit assignment and prioritizes task resolution over token efficiency through hierarchical policy optimization.
Extensive experiments on SWE-bench series demonstrate that HERO significantly outperforms existing methods and achieves a favorable trade-off between resolution rate and token efficiency.
In the future, we plan to extend HERO to more complex long-horizon agentic tasks, such as scientific data analysis and engineering design.



\bibliography{references}
\bibliographystyle{preprint}

\clearpage
\appendix
\begin{leftline}
    {\LARGE\textsc{Appendix}}
\end{leftline}
\etocdepthtag.toc{mtappendix}
\etocsettagdepth{mtchapter}{none}
\etocsettagdepth{mtappendix}{subsection}
{
    \footnotesize
    \etocsettocstyle{}{}
    \tableofcontents
}

\newpage
\section{More Implementation Details}
\label{sec:implementation-details}
In this section, we provide more implementation details about the training data, implementation settings, entropy estimation, and identification of unproductive behaviors.

\subsection{Training Data}
\label{sec:training-data-details}
As mentioned in Section~\ref{sec:experiments}, we employ 640 multilingual SWE tasks for training.
Specifically, the training set contains 352 Python tasks, 38 Java tasks, and 50 tasks each in C/C++, JavaScript/TypeScript, PHP, Go, and Rust.
Such a collection covers SWE tasks across multiple programming languages, endowing the training of coding agents with multilingual reasoning capabilities.

\subsection{More Implementation Settings}
\label{sec:scaffold-evaluation-details}
We provide additional implementation details for training and evaluation.
Specifically, the default tool configuration disables WebFetch, WebSearch, and Agent, while retaining all other scaffold-provided tools.
For training, each training rollout batch contains $8$ tasks with $16$ sampled trajectories per task, yielding $128$ trajectories in total.
Moreover, in Eq.~\ref{eq: hierarchical-rl}, the derived turn-level advantage is assigned to every generated token within the turn.
Note that, given the stronger issue-solving capability of Qwen3.5-35B-A3B, the trade-off coefficient $\lambda$ and threshold $\tau$ in Eq.~\ref{eq: adaptive-weight} are set to $0.1$ and $0.8$, respectively.
For evaluation, we use the checkpoint from the final training iteration.
Together, the default settings are summarized in Table~\ref{tab:evaluation-settings}.

\begin{table}[H]
\centering
\caption{Default implementation settings.}
\label{tab:evaluation-settings}
\tablestyle{12pt}{1.3}
\begin{tabular}{llc}
\toprule
Category & Parameter & Setting \\
\midrule
Agent Scaffold & Claude Code version & 2.1.111 \\
\midrule
Training & Tasks per rollout batch & 8 \\
         & Rollout group size & 16 \\
         & Trajectories per rollout batch & 128 \\
\midrule
Evaluation & Sampling temperature & 0.6 \\
           & Top-$p$ & 0.95 \\
           & Maximum interaction turns & 200 \\
           & Context length & 131,072 tokens \\
\bottomrule
\end{tabular}
\end{table}

\subsection{Entropy Estimation}
\label{sec:entropy-estimation-details}
We provide additional details on the estimation of turn-level entropy.
Following~\citet{top10,rtwi}, we estimate token-level entropy using the top-$10$ token probabilities and derive turn-level entropy as
\begin{equation}
    h_{i,j}
    =
    \frac{1}{|\mathcal{T}_{i,j}|}
    \sum_{k\in\mathcal{T}_{i,j}} H_k,
\end{equation}
where $\mathcal{T}_{i,j}$ denotes the set of generated tokens in the $j$-th turn of rollout $i$, with $H_k$ denotes the entropy of the $k$-th generated token.

\subsection{Identification of Unproductive Behaviors}
\label{sec:unproductive-behavior-details}
In the manuscript, we analyze the association between turn-level entropy and unproductive behaviors in Fig.~\ref{fig: motivation}.
Here, we provide the definition for repeated tool calls, excessive exploration, and erroneous commands in Table~\ref{tab:behavior-identification}.
Note that each behavior is counted at most once per turn. 

\begin{table}[H]
\centering
\caption{Identification rules for unproductive behaviors.}
\label{tab:behavior-identification}
\tablestyle{5pt}{1.3}
\begin{tabular}{p{0.22\linewidth}p{0.71\linewidth}}
\toprule
Behavior & Identification Rule \\
\midrule
Repetition &
Repeats the same sequence of tool calls and arguments as the preceding turn. \\
\addlinespace
Overexploration &
Uses only \texttt{Glob}, \texttt{Grep}, \texttt{Read}, or \texttt{NotebookRead} for at least 15 consecutive turns. \\
\addlinespace
Tool Error &
Tool responses contain explicit tool-error markers, excluding failed tests or nonzero exit codes alone. \\
\addlinespace
Any Issue & Exhibits at least one of the above behaviors. \\
\bottomrule
\end{tabular}
\end{table}

\newpage

\section{Additional Experimental Results}
\label{sec: additional-experiments}
In this section, we present more experimental results of the proposed HERO. Unless otherwise specified, all experiments are conducted using Qwen3.5-4B.

\subsection{Comparisons with Token-Efficient Reinforcement
Learning Methods}
\label{sec:training-efficient-comparison}
As discussed in Introduction, existing token-efficient RL methods are specifically designed for non-interactive reasoning or simple agentic tasks, which struggle to achieve a favorable trade-off between resolution rate and token efficiency in complex SWE scenarios.
We compare HERO with four state-of-the-art token-efficient RL methods, \textit{i.e.}, Training Efficient~\citep{efficient_rl1}, LASER~\citep{LASER}, OTC-GRPO~\citep{otc}, and SlimSearcher~\citep{SlimSearcher}, on SWE-bench Verified and SWE-bench Multilingual.
For fair comparison, we adopt the same training and evaluation settings across methods, except for method-specific configurations.
Specifically, for Training Efficient, OTC-GRPO, and SlimSearcher, we follow the configurations reported in their respective papers. 
While for LASER, we use the fixed-threshold variant with a threshold of 8,192 total output tokens across interaction turns, estimated based on the generation lengths typical of SWE tasks.

From the results in Table~\ref{tab:training_efficient_4b}, one could have the following observations and conclusions.
On the one hand, HERO achieves higher resolution rates with lower token usage than Training Efficient and SlimSearcher on both benchmarks.
On the other hand, although LASER and OTC-GRPO consume fewer tokens, these token savings come at the cost of degrading resolution rates, highlighting the difficulty of balancing resolution performance and token efficiency for existing token-efficient RL methods.
Overall, HERO significantly outperforms existing token-efficient RL methods in the TRS metric, demonstrating that HERO achieves a more favorable trade-off between the two objectives.

\begin{table}[H]
\centering
\tablestyle{5pt}{1.15}
\caption{Comparisons with token-efficient RL methods on SWE-bench Verified and SWE-bench Multilingual.}
\label{tab:training_efficient_4b}
\begin{tabular}{lcccccc}
\toprule
& \multicolumn{3}{c}{SWE-bench Verified}
& \multicolumn{3}{c}{SWE-bench Multilingual} \\
\cmidrule(lr){2-4} \cmidrule(lr){5-7}
Method
& Resolved Rate (\%) & Tokens (M) & TRS
& Resolved Rate (\%) & Tokens (M) & TRS \\
\midrule
Base              & 32.8 & 4.5 & 14.0 & 20.0 & 4.0 & 8.9 \\
GRPO              & 37.4 & 6.8 & 13.4 & 22.0 & 7.6 & 7.5 \\
Training Efficient& 37.4 & 6.4 & 13.7 & 21.3 & 6.8 & 7.6 \\
LASER             & 33.2 & 3.4 & 15.8 & 21.7 & 4.8 & 9.0 \\
OTC-GRPO          & 32.4 & 4.0 & 14.5 & 17.0 & 3.2 & 8.3 \\
SlimSearcher      & 37.6 & 6.3 & 13.9 & 21.0 & 6.4 & 7.7 \\
\rowcolor{pink!30}
Ours              & 40.0 & 4.5 & 17.1 & 24.3 & 4.5 & 10.4 \\
\bottomrule
\end{tabular}
\end{table}

\subsection{Comparisons with Inference-Time Methods}
\label{sec:inference-time-comparison}
In the manuscript, we argue that existing inference-time approaches struggle to balance the trade-off between resolution performance and token efficiency.
To verify the claim, we conduct additional experiments on four representative inference-time approaches, \textit{i.e.}, SWE-Pruner~\citep{swe_pruner}, Pruning + Summ.~\citep{compact2}, Observation Masking~\citep{complexity_trap}, and Turn Limit~\citep{turn_limit}.
Specifically, we apply inference-time approaches to both Qwen3.5-4B and its HERO-trained counterpart, termed HERO-4B. 

From the results in Table~\ref{tab:inference-time-performance}, one could have the following observations and conclusions:
i) most inference-time approaches reduce token usage without improving resolution performance over the base model, highlighting the difficulty of balancing resolution performance and token efficiency; 
ii) HERO-4B achieves higher TRS scores than inference-time approaches applied to the base model, suggesting that our training-time designs achieve a more favorable trade-off between resolution performance and token efficiency;
iii) thanks to the training-time designs, HERO-4B could still benefit from inference-time approaches, suggesting that the proposed HERO could complement inference-time approaches.

\begin{table}[H]
\centering
\tablestyle{5pt}{1.15}
\caption{Comparisons with inference-time approaches on SWE-bench Verified and SWE-bench Multilingual.}
\label{tab:inference-time-performance}
\begin{tabular}{lcccccc}
\toprule
& \multicolumn{3}{c}{SWE-bench Verified}
& \multicolumn{3}{c}{SWE-bench Multilingual} \\
\cmidrule(lr){2-4} \cmidrule(lr){5-7}
Variants
& Resolved Rate & Tokens & TRS
& Resolved Rate & Tokens & TRS \\
\midrule
Qwen3.5-4B  
& 32.8 & 4.5 & 14.0 & 20.0 & 4.0 & 8.9 \\
\hspace{1.0em} + SWE-Pruner
& 33.0 & 4.3 & 14.3 & 19.7 & 3.8 & 9.0 \\
\hspace{1.0em} + Pruning + Summ.
& 31.4 & 3.2 & 15.3 & 18.7 & 3.3 & 9.0 \\
\hspace{1.0em} + Observation Masking
& 30.0 & 3.8 & 13.7 & 18.0 & 3.5 & 8.5 \\
\hspace{1.0em} + Turn Limit
& 29.0 & 3.0 & 14.5 & 16.7 & 3.1 & 8.2 \\
\midrule
HERO-4B
& 40.0 & 4.5 & 17.1 & 24.3 & 4.5 & 10.4 \\
\hspace{1.0em} + SWE-Pruner
& 40.4 & 4.3 & 17.5 & 23.7 & 4.4 & 10.2 \\
\hspace{1.0em} + Pruning + Summ.
& 36.4 & 3.4 & 17.4 & 21.7 & 3.6 & 10.1 \\
\hspace{1.0em} + Observation Masking
& 39.4 & 3.8 & 18.0 & 23.4 & 3.8 & 10.7 \\
\hspace{1.0em} + Turn Limit
& 36.0 & 3.1 & 17.8 & 21.0 & 3.1 & 10.4 \\
\bottomrule
\end{tabular}
\end{table}

\subsection{Analytic Study on Prefix-Cache Reuse and Inference Cost}
\label{sec:inference-time-cache}
To further investigate efficiency in real-world applications, we conduct an analytic study on prefix-cache reuse and relative inference cost on SWE-bench Verified.
Following~\citet{complexity_trap}, we estimate inference cost post hoc from recorded token usage.
Specifically, we derive the token-weighted cache-hit rate and per-instance cost as
\begin{equation}
\begin{gathered}
    \mathrm{HitRate}
    =
    100\frac{\sum_i t_i^{\mathrm{cache}}}
    {\sum_i t_i^{\mathrm{in}}},\\
    \mathrm{Cost}_i
    =
    \left(t_i^{\mathrm{in}}-t_i^{\mathrm{cache}}\right)
    +0.1t_i^{\mathrm{cache}}
    +4t_i^{\mathrm{out}},
\end{gathered}
\end{equation}
where $t_i^{\mathrm{in}}$, $t_i^{\mathrm{cache}}$, and $t_i^{\mathrm{out}}$ denote the cumulative input, cached input, and output token counts for instance $i$.
We employ fixed cost weights of $1$, $0.1$, and $4$ to uncached input, cached input, and output tokens, respectively.
Relative inference cost is normalized by the average per-instance cost of the corresponding base model at each model size.

From the results in Table~\ref{tab:inference-time-cache}, one could have the following conclusions.
On the one hand, although context manipulation methods, \textit{e.g.}, Pruning + Summ. and Observation Masking, could reduce token usage, these methods would undermine prefix-cache reuse, resulting in higher relative inference costs.
On the other hand, interaction limits reduce both token usage and relative inference cost at the expense of a serious degradation in resolution performance.
In contrast, HERO achieves competitive prefix-cache reuse and relative inference cost, supporting its effectiveness in practical applications.

\begin{table}[h]
\centering
\tablestyle{5pt}{1.2}
\caption{Analytic study on prefix-cache reuse and relative inference cost on SWE-bench Verified.}
\label{tab:inference-time-cache}
\begin{tabular}{lccc}
\toprule
Variants & Resolved Rate (\%) & Cache Hit (\%) & Relative Cost \\
\midrule
% \multicolumn{4}{c}{\textit{\textbf{4B Size}}} \\
% \midrule
Qwen3.5-4B & 32.8 & 92.4 & 1.0 \\
\hspace{1.0em} + SWE-Pruner & 33.0 & 86.3 & 1.4 \\
\hspace{1.0em} + Pruning + Summ. & 31.4 & 71.9 & 1.3 \\
\hspace{1.0em} + Observation Masking & 30.0 & 80.3 & 1.3 \\
\hspace{1.0em} + Turn Limit & 29.0 & 89.2 & 0.7 \\
\rowcolor{pink!30}
HERO-4B & 40.0 & 91.9 & 1.0 \\
% \midrule
% \multicolumn{4}{c}{\textit{\textbf{9B Size}}} \\
% \midrule
% Qwen3.5-9B & 45.0 & 90.6 & 1.0 \\
% \hspace{1.0em} + SWE-Pruner & -- & -- & -- \\
% \hspace{1.0em} + Pruning + Summ. & -- & -- & -- \\
% \hspace{1.0em} + Observation Masking & -- & -- & -- \\
% \hspace{1.0em} + Turn Limit & -- & -- & -- \\
% \rowcolor{pink!30}
% HERO-9B & 49.2 & 91.0 & 1.0 \\
\bottomrule
\end{tabular}
\end{table}

\subsection{Ablation Studies on SWE-bench Multilingual}
\label{sec: multilingual-ablation}
In the manuscript, we have conducted ablation studies on SWE-bench Verified. 
Here, we extend the ablation studies to more complex multilingual SWE tasks, \textit{i.e.}, SWE-bench Multilingual.
As shown in Table~\ref{tab:ablation_multilingual}, removing hierarchical policy optimization (HPO) reduces token usage at the cost of degrading resolution performance, while either coarse-grained or fine-grained efficiency advantages contribute to reducing token consumption and improving resolution performance.

\begin{table}[H]
\centering
\caption{Ablation studies of HERO on SWE-bench Multilingual.}
\label{tab:ablation_multilingual}
\tablestyle{8pt}{1.15}
\begin{tabular}{lccc}
\toprule
Variants & Resolved Rate & Tokens & TRS \\
\midrule
% Baseline   & 20.0 & 4.0 & 8.9 \\
GRPO     & 22.0 & 7.6 & 7.5 \\
w/o HPO     & 17.0 & 2.3 & 9.4 \\
w/o Coarse & 23.7 & 7.0 & 8.4 \\
w/o Fine   & 22.3 & 5.5 & 8.7 \\
\midrule
Default    & 24.3 & 4.5 & 10.4 \\
\bottomrule
\end{tabular}
\end{table}

\subsection{Analytic Study on Token Usage Breakdown}
We conduct more in-depth analysis to investigate the overhead of multi-turn interactions.
From the results in Tables~\ref{tab:token_breakdown} and~\ref{tab:token_breakdown_multilingual}, input tokens account for more than 99\% of total consumption, highlighting the significant overhead of repeatedly processing accumulated context.
Compared with GRPO, HERO achieves higher resolution rates with fewer interaction turns and lower token consumption on both benchmarks.

\begin{table}[H]
\centering
\tablestyle{3pt}{1.15}
\caption{Token usage and interaction turns on SWE-bench Verified. ``Input Ratio'' denotes the proportion of input tokens in total token consumption.}
\label{tab:token_breakdown}
\begin{tabular}{lcccc}
\toprule
\textbf{Method}
& \textbf{Resolved Rate}
& \textbf{Total Tokens}
& \textbf{Input Ratio}
& \textbf{Avg. Turns} \\
\midrule
Qwen3.5-4B
& 32.8 & 4.5 & 99.6 & 68.7 \\
\hspace{1.0em} + GRPO
& 37.4 & 6.8 & 99.6 & 94.8 \\
% \rowcolor{pink!30}
\hspace{1.0em} + Ours
& 40.0 & 4.5 & 99.4 & 65.6 \\
\bottomrule
\end{tabular}
\end{table}

\begin{table}[H]
\centering
\tablestyle{3pt}{1.15}
\caption{Token usage and interaction turns on SWE-bench Multilingual.}
\label{tab:token_breakdown_multilingual}
\begin{tabular}{lcccc}
\toprule
\textbf{Method}
& \textbf{Resolved Rate}
& \textbf{Total Tokens}
& \textbf{Input Ratio}
& \textbf{Avg. Turns} \\
\midrule
Qwen3.5-4B
& 20.0 & 4.0 & 99.5 & 64.3 \\
\hspace{1.0em} + GRPO
& 22.0 & 7.6 & 99.6 & 106.5 \\
% \rowcolor{pink!30}
\hspace{1.0em} + Ours
& 24.3 & 4.5 & 99.3 & 69.6 \\
\bottomrule
\end{tabular}
\end{table}

\subsection{Analytic Study on Scaffold Generalization}
In the manuscript, we have evaluated the proposed HERO under the unified Claude Code scaffold for fairness. 
Here, we conduct additional experiments using widely-used mini-SWE-agent scaffold on SWE-bench Verified.
As shown in Table~\ref{tab:different_scaffolds}, even training on the Claude Code scaffold, HERO achieves higher resolution rates with lower token consumption than GRPO under both scaffolds, indicating HERO learns efficient reasoning patterns that generalize across scaffolds.

\begin{table}[H]
\centering
\tablestyle{6pt}{1.15}
\caption{Performance comparisons under different agent scaffolds on SWE-bench Verified.}
\label{tab:different_scaffolds}
\begin{tabular}{lcccccc}
\toprule
& \multicolumn{3}{c}{Claude Code}
& \multicolumn{3}{c}{mini-SWE-agent} \\
\cmidrule(lr){2-4} \cmidrule(lr){5-7}
Variants
& Resolved Rate & Tokens & TRS
& Resolved Rate & Tokens & TRS \\
\midrule
Qwen3.5-4B
& 32.8 & 4.5 & 14.0
& 39.5 & 4.0 & 17.7 \\
\hspace{1.0em} + GRPO
& 37.4 & 6.8 & 13.4
& 41.2 & 5.2 & 16.5 \\
\rowcolor{pink!30}
\hspace{1.0em} + Our
& 40.0 & 4.5 & 17.1
& 43.0 & 4.0 & 19.2 \\
\bottomrule
\end{tabular}
\end{table}

\subsection{More Detailed Ablation Studies}
\label{sec:hrl-outcome-ablation}
To further verify the effectiveness of individual designs, we carry out ablation studies on the efficiency gating and resolution-first clipping in hierarchical policy optimization (HPO) and outcome conditioning in coarse-grained efficiency advantage.
As demonstrated in Table~\ref{tab:detailed_ablation}, one could have the following conclusions:
i) either efficiency gating or resolution-first clipping could help preserve resolution performance, which verifies the effectiveness of hierarchical policy optimization in prioritizing task resolution over token efficiency;
ii) resolution-calibrated credit assignment could facilitate the
learning of efficient reasoning patterns without conflating token efficiency with resolution performance, thus achieving a better trade-off between resolution rate and token efficiency.

\begin{table}[H]
\centering
\tablestyle{3pt}{1.15}
\caption{More detailed ablation studies on SWE-bench Verified. ``w/o Calibration'' denotes using shared threshold across all trajectories within the rollout group.}
\label{tab:detailed_ablation}
\begin{tabular}{llccc}
\toprule
Component & Variants & Resolved Rate & Tokens & TRS \\
\midrule
\multirow{3}{*}{HPO}
& w/o Efficiency Gating & 36.0 & 3.8 & 16.4 \\
& w/o Clipping & 37.8 & 4.1 & 16.7 \\
& Default & 40.0 & 4.5 & 17.1 \\
\midrule
\multirow{2}{*}{Coarse}
& w/o Calibration & 38.2 & 5.9 & 14.5 \\
& Default & 40.0 & 4.5 & 17.1 \\
\bottomrule
\end{tabular}
\end{table}

\subsection{Sensitivity to the Capability Threshold}
\label{sec: capability-threshold}
\begin{figure}[H]
    \centering
    \includegraphics[width=0.4\linewidth]{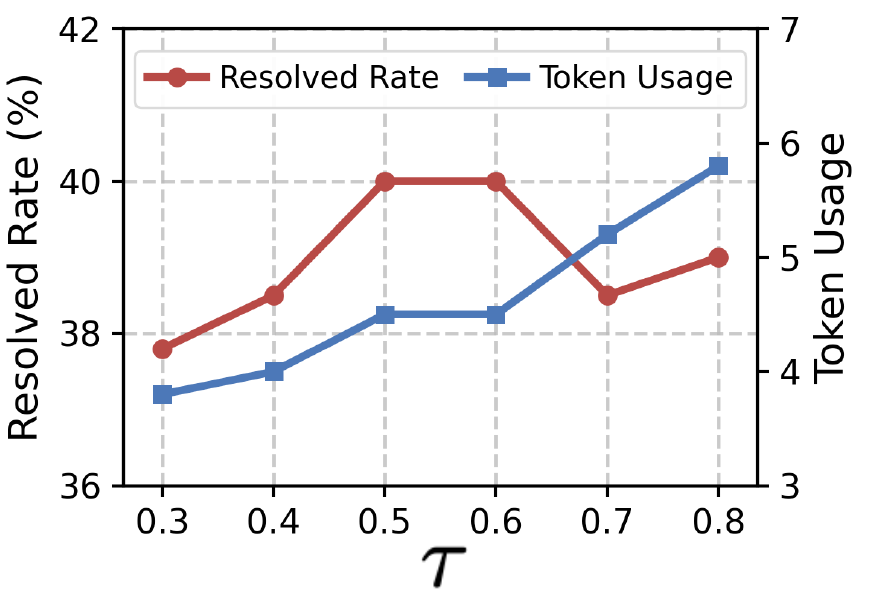}
    \caption{Parameter analysis of the capability threshold $\tau$ in Eq.~\ref{eq: adaptive-weight}.}
    \label{fig: parameter-tau}
\end{figure}

We further investigate the effect of the capability threshold $\tau$ in Eq.~\ref{eq: adaptive-weight}.
As shown in Fig.~\ref{fig: parameter-tau}, HERO achieves a desirable trade-off between resolution performance and token efficiency when $\tau\in[0.5,0.6]$.

\subsection{Sensitivity to the Entropy Percentile}
\label{sec: entropy-percentile}
\begin{figure}[H]
    \centering
    \includegraphics[width=0.4\linewidth]{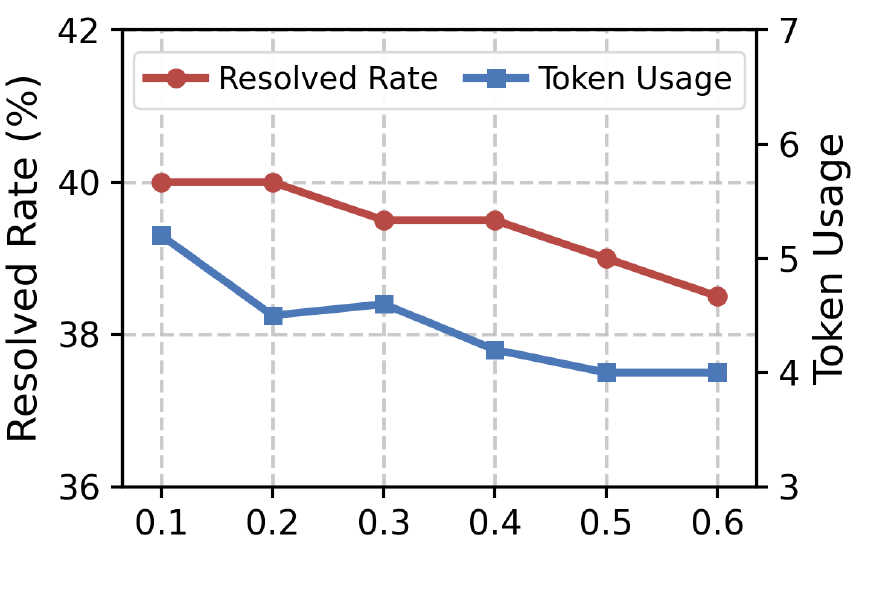}
    \caption{Parameter analysis of the entropy percentile in Eq.~\ref{eq: fine-advantage}.}
    \label{fig: parameter-sup}
\end{figure}

We conduct additional parameter analysis of the entropy percentile in Eq.~\ref{eq: fine-advantage}.
As shown in Fig.~\ref{fig: parameter-sup}, a larger percentile generally reduces token usage but also degrades resolution performance, suggesting that penalizing too many turns might suppress necessary reasoning.

\subsection{Analytic Study on Fine-grained Turn Advantage}
\label{sec:turn-selection-strategies}
To further investigate the effectiveness of the fine-grained turn advantage, we compare two alternative penalty strategies on SWE-bench Verified: i) Random: randomly selecting 20\% of turns for penalization; ii) High Entropy: selecting the 20\% of turns with the highest entropy for penalization.
As demonstrated in Table~\ref{tab:turn-selection-strategies}, the default strategy of penalizing low-entropy turns achieves a higher resolution rate with lower token consumption than both alternatives, suggesting that penalizing low-entropy turns contributes to learning efficient reasoning patterns.

\begin{table}[H]
\centering
\caption{Analytic study of turn-level penalty strategies on SWE-bench Verified.}
\label{tab:turn-selection-strategies}
\tablestyle{8pt}{1.15}
\begin{tabular}{lccc}
\toprule
Variants & Resolved Rate (\%) & Tokens (M) & TRS \\
\midrule
Random & 36.8 & 5.4 & 14.5 \\
High Entropy & 36.0 & 6.0 & 13.6 \\
\midrule
Default & 40.0 & 4.5 & 17.1 \\
\bottomrule
\end{tabular}
\end{table}

\end{document}